# Anomalous lattice thermal conductivity in layered materials MNCl (M=Zr, Hf) driven by the lanthanide contraction

Xiaoxia Yu[a], Hezhu Shao[b], Xueyun Wang[a], Yingcai Zhu[a], Daining Fang[c, d], Jiawang Hong[a, *]

[a] School of Aerospace Engineering, Beijing Institute of Technology, Beijing 100081, China

[b] College of Electrical and Electronic Engineering, Wenzhou University, Wenzhou 325035, China

[c] Institute of Advanced Structure Technology, Beijing Institute of Technology, Beijing 100081, China

[d] College of Engineering, Peking University, Beijing 100871, China

**Corresponding Author**

*E-mail: hongjw@bit.edu.cn.

**Abstract:** High performance thermoelectric devices requires materials with low lattice thermal conductivities. Many strategies, such as phonon engineering, have been made to reduce lattice thermal conductivity without simultaneously decrease of the charge transport performance. It is a simple and effective approach to use materials with heavy element to reduce the lattice thermal conductivity. Here, based on the first-principles calculations and phonon Boltzmann transport equations, we find the replacement of Zr with heavy element Hf in ZrNCl doesn't reduce the lattice thermal conductivity, instead, it surprisingly increases by about 4 times at 300K. This unusual lattice thermal conductivity is mainly attributed to the dramatic enhancement in phonon lifetimes in Hf compound, originating from the strong interatomic bonding due to lanthanide contraction. Our findings unveil the microscopic mechanisms of high thermal transport properties in materials with heavy element, providing an alternative strategy in materials design with low lattice thermal conductivity for thermoelectric applications such as power restoration and generation.

I. INTRODUCTION

Thermoelectric (TE) materials, which directly transfer heat energy to electricity, are critical for cooling and power generating applications to solve the energy crisis and environmental issues.[1] The performance of TE materials is characterized by dimensionless figure-of-merit $ZT = S^2\sigma T/(\kappa_e+\kappa_{latt})$, where $S$, $\sigma$, $\kappa_e$ and $\kappa_{latt}$ are Seebeck coefficient, electrical conductivity, electrical and lattice thermal conductivity, respectively.[2] Since $S$, $\sigma$ and $\kappa_e$ strongly couples with each other, optimization of thermoelectric property by tuning one single parameter is an difficult task.[3] However, reducing the lattice thermal conductivity ($\kappa_{latt}$) is an effective strategy to boost thermoelectric performance as it's a relatively independent parameter.[4] An intrinsically low $\kappa_{latt}$[5–9] is expected in materials with features such as complex crystal structures[10,11], liquid-like ions[12] and strong anharmonicity[13] etc.

Lattice thermal conductivity is sensitive to the constituent atoms mass. Usually, the heavy atomic masses gives rise to low acoustic phonon frequency, low group velocity and depressed lattice thermal conductivity.[14–18] For example, PbTe shows the lattice thermal conductivity as 2.3 W/mK at 300 K with mean sound velocity about 1.8 km/s, while the lighter element compounds, i.e., PbSe and PbS, they show higher $\kappa_{latt}$ of 2.64 and 2.80 W/mK, with higher mean sound velocity of 1.9 and 2.1 km/s, respectively.[19] As a consequence, substituting with heavy mass element is an effective strategy to reduce lattice thermal conductivity and improve the figure of merit of thermoelectrics.[20] However, it is not always true that materials with heavy element exhibit low lattice thermal conductivity than those with light element. Because the lattice thermal conductivity depends on not only the phonon group velocity (related to the mass) but also the phonon relaxation time (related to the anharmonicity and scattering space) For instance, a high $\kappa_{latt}$ was observed in PbSe despite of its larger atomic mass than that in SnSe, which was attributed to the giant phonon scattering originating from an unstable electronic structure with orbital interactions, leading to a ferroelectric-like lattice instability in SnSe[13].

In this work, we find that the replacement of Zr with heavy element Hf in ZrNCl surprisingly enhances the thermal conductivity by about 4 times at 300K, rather than reduces it due to the lanthanide contraction effect. Transition metal nitride halides MNCl (M=Zr, Hf) family was first reported in 1964 and then widely applied as superconducting materials for a long time. Crystals of MNCl[21], β-type SmSI structure, have relatively lower formation energies.[22] The layered transition metal nitrides exhibited diverse good superconducting properties[23,24] and the monolayer p-type ZrNCl recently was predicted to show good electronic transport properties and high figure-of-merit.[25] However, the thermal transport properties of MNCl is barely discussed in contrast to their electric properties. Here, we investigate the lattice thermal transport for MNCl (M=Zr, Hf) based on the first-principles calculations and phonons Boltzmann transport equations. It shows that HfNCl possesses larger lattice thermal conductivity than that of ZrNCl, due to the lanthanide contraction effect. Though heavy mass of Hf induces low group velocity and tends to reduce the thermal conductivity, the stronger bonds in HfNCl broaden the phonon gap and reduces the phonon scattering phase space. This suppresses significantly the three-phonon scattering channels and overcomes the reduction effect of heavy mass, therefore, it enhances the lattice thermal conductivity of HfNCl dramatically. Our findings provide an alternative strategy in designing materials with low lattice thermal conductivity for thermoelectric related applications.

## II. COMPUTATIONAL DETAILS

First-principles calculations are carried out by using the plane wave pseudopotential approach as implemented in the Vienna ab initio simulation package (VASP)[26]. In order to describe the exchange–correlation effect, generalized gradient approximation (GGA) with the Perdew–Burke-Ernzerhof (PBE) exchange–correlation potential is employed[27]. The van der Waals density functional method of optB86b vdW-DF is used to treat dispersion interactions for better description of interlayer interactions[28]. For structural relaxation, a kinetic-energy cutoff for the plane-wave expansion is set to 500 eV, all the atoms in the primitive cell are fully relaxed until the force on each atom is less than 0.001 eV/Å. Electronic minimization was performed with a tolerance of $10^{-6}$ eV and the Brillouin zones are sampled with 8×8×8 Monkhorst-Pack $k$-point meshes[29]. For the lattice thermal conductivity simulation, harmonic and anharmonic force constants are required for phonon Boltzmann transport equation[30], which is performed in Phonopy package[31] and ShengBTE code[32], respectively. The lattice thermal conductivity is calculated with 2×2×2 the supercell. To gain insights into the chemical bonding, we perform the crystal orbital Hamilton population (COHP) analysis for different pairs of atoms within the crystal structure utilizing LOBSTER code[33].

## III. RESULTS AND DISCUSSION

A conventional crystal of layered MNCl includes three (MNCl)$_2$ slabs as shown in Fig. 1, where there are three kinds of elements and consist of 18 atoms in total with space group *R-3m* (No.166). Equilibrium crystallographic parameters with the corresponding available literature data for the conventional cell are listed in Table 1. It can be seen that our first-principles results are in well agreement with other theoretical[34] and experimental results[35]. Note that all bond lengths connected with Hf atom are smaller than those of Zr atom, which are consistent with the experimental measurements (detailed discussions later). Strong anisotropy is expected in this system due to weak van der Waals interaction between two slabs and strong chemical in-plane bonds.

Phonon dispersion and phonon density of states (PDOS) of MNCl are shown in Fig. 2. The Raman results of ZrNCl are also plotted in Fig. 2(a) and our first-principles results agrees well with experiment data[36]. From Fig. 2, we can see that both crystals have similar phonon dispersive shape, for example, a phonon gap appears due to large mass discrepancy[37], and the phonon cures along Γ–Z direction are less dispersive attributed to weak van der Waals interaction [38]. The phonon gap of ZrNCl is14 meV and HfNCl of 20 meV. The larger phonon gap in Hf compound is due to upward shift of high frequency phonons and downward shift of low frequency phonons simultaneously as shown in Fig. 2(b), which depends on the interatomic force and component atomic mass. From PDOS it can be clearly seen that the low frequency phonons mainly from heavier Zr/Hf and Cl atoms, while the lighter N atoms are the only contributors for the high frequency optical phonons modes. The analysis of PDOS reveals low frequency phonon branches originating from the collected vibration of M and Cl atoms are strongly hybridized below 40 meV.

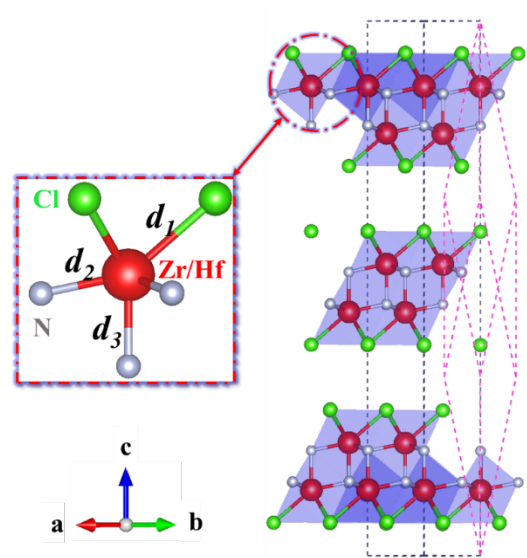

Fig. 1. Schematic of crystal lattice of MNCl, (M=Zr, Hf). Green, red and gray spheres represent Cl, Zr/Hf and N atoms, respectively. Gray and pink dashed line represent conventional and primitive cell respectively. Bonds $d_1$, $d_2$ and $d_3$ indicate the bond of M-Cl, M-N in-plane and M-N out-of-plane as labeled in zoom-in inset.

We calculate the lattice thermal conductivities of ZrNCl and HfNC as a function of temperature, as shown in Fig. 3(a) and (b). It can be seen that MNCl shows strong anisotropy in thermal conductivity due to its layered structure. For example, the $\kappa_{latt}$ along $c$-axis of ZrNCl is an order of magnitude smaller than that along $a$-axis. Interestingly, the results show that HfNCl exhibits higher $\kappa_{latt}$ along both $a$ and $c$-axis than that of ZrNCl containing light element, which is contradictory to the conventional view that the heavy crystals usually possess low lattice thermal conductivity[39,40]. For example, the in-plane thermal conductivity of HfNCl is 9.90 W/mK at 300K, which is nearly four times than that of ZrNCl (2.56 W/mK). The frequency dependence of cumulative in-plane $\kappa_{latt}$ is shown in Fig. 3(c). It can be seen that the curves for both compounds increase sharply below ~30 meV and then keep almost constant, though there is a slight increase around ~60 meV for HfNCl. This indicates that the acoustic phonons and low energy optical phonons below 30 meV dominate the contribution (about 95%) to the $\kappa_{latt}$ in the layered MNCl compounds.

Table 1 Structural and mass parameters of the conventional cell in comparison with experimental results[35].

| Formula | $a$ (Å) | $c$ (Å) | $d_1$ (Å) | $d_2$ (Å) | $d_3$ (Å) | $m_{MCl}/m_N$ |
|---|---|---|---|---|---|---|
| ZrNCl | 3.642 (3.605[a]) | 27.679 (27.672[a]) | 2.772 | 2.152 (2.130[a]) | 2.197 (2.172[a]) | 9.045 |
| HfNCl | 3.581 (3.576[a]) | 27.621 (27.711[a]) | 2.744 | 2.116 (2.114[a]) | 2.167 (2.163[a]) | 15.273 |

[a]Reference [35].

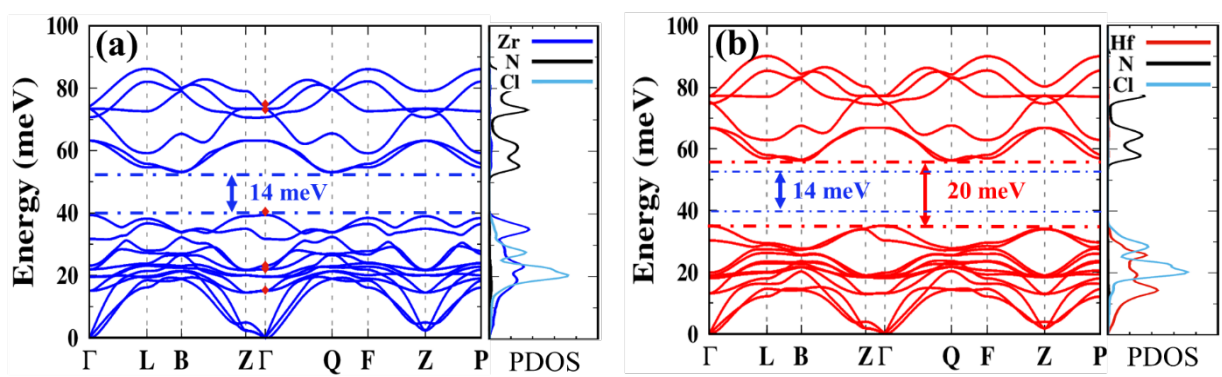

Fig. 2. Phonon dispersion and phonon partial density of states (PDOS) of (a) ZrNCl and (b) HfNCl. Red diamond in (a) is Raman data.[36] The phonon band gap of ZrNCl (blue dashed line) is 14 meV and 20 meV for Hf (red dashed line). The phonon gap of ZrNCl is also shown in blue in Fig. 2(b) for comparisons.

To reveal the mechanism of this unusual phenomena that large mass HfNCl possesses lower lattice thermal conductivity than small mass ZrNCl compound, we firstly calculate the group velocities of acoustic phonon modes along *a* direction for both materials, as listed in Table 2. It shows that the group velocities of HfNCl is smaller than that of ZrNCl, which is in consistent with conditional view and can't explain their unusual lattice thermal conductivities. Then we focus on the phonon scattering rate $\tau_{\mathbf{q}s}$ since the lattice thermal conductivity $\kappa_{latt}$ depends on not only group velocity $v_{\mathbf{q}s}$, but also phonon scattering rate and heat capacity $C_{\mathbf{q}s}$,[13]

$$\kappa_{latt} = \frac{1}{3}\sum_{\mathbf{q}s} C_{\mathbf{q}s} v_{\mathbf{q}s}^2 \tau_{\mathbf{q}s} \qquad (1)$$

where **q** is the phonon branches of wave vector and *s* is branch index.

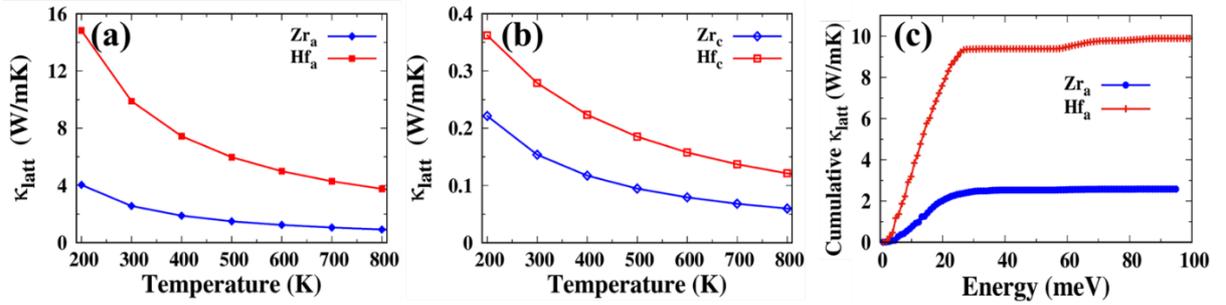

Fig. 3. Temperature dependence of lattice thermal conductivity of ZrNCl and HfNCl along (a) *a*- and (b) *c*-axis, (c) frequency dependence of cumulative in-plane lattice thermal conductivity at room temperature.

We calculate the phonon-phonon scattering rates and scattering phase space, as shown in Fig. 4. It can be seen that the phonon scattering rates of ZrNCl is higher than that of HfNCl in the whole energy range, suggesting a depressed $\kappa_{latt}$ in Zr compound. In particular, the phonon scattering rates of ZrNCl are significantly larger than that of HfNCl with frequency above 50 meV, implying the high energy optical phonons in Zr compound have higher probability to scatter phonons. This can be clearly seen from Fig. 4(b) in which the phonon-phonon scattering phase space of aoo and ooo processes for ZrNCl ('a' and 'o' indicate acoustic and optical phonon, respectively) are larger than that of HfNCl. The reason for this reduced phase space in HfNCl is that the scattering channels are affected by large phonon gap between low-energy and high-energy optical phonons due to large mass ratio, as shown in Fig. 2(b). With such a large phonon gap, the low-energy acoustic and optical phonons below the gap will have fewer chances to scatter the elevated high-energy optical modes above the gap, causing smaller (aoo) and (ooo) three-phonon scattering space and hence small scattering rates in HfNCl. Therefore, though large mass induces small group velocity in HfNCl, its small phonon scattering rates due to large phonon gap overcome the effect of reduced group velocity and cause the high lattice thermal conductivity, compared with ZrNCl.

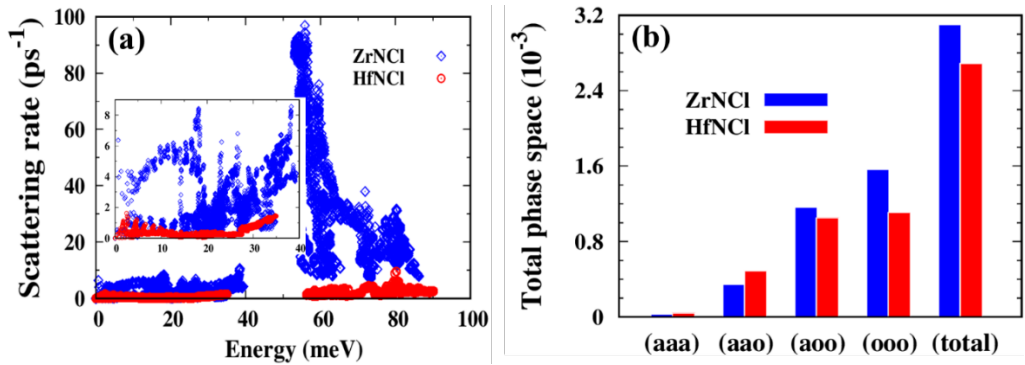

Fig. 4. (a) Frequency dependence of three-phonon scattering rates, (b) phonon scattering phase space. The temperate is 300 K. The three-phonon scattering processes are labeled as (aaa), (aao), (aoo) and (ooo), where 'a' and 'o' indicate acoustic and optical phonons in three-phonon scattering processes.

We can see that the large phonon gap in HfNCl plays a critical role on its high thermal conductivity. The large phonon gap results from the upward shift of high frequency phonons and downward shift of low frequency phonons simultaneously as shown in Fig. 2(b). It can be easily understood that the low-energy acoustic and optical phonons move downward due to the large mass of Hf, compared with ZrNCl. However, it seems a little strange that the high-energy optical phonons beyond the gap, which are contributed purely from N (as indicated from PDOS in Fig. 2), move upward in HfNCl. As is known, usually, the heavy element in the same group (here, Hf vs Zr) will have large radius and hence weak bonding to other atoms (here N). Therefore, optical phonons from N should move downward rather than upward in HfNCl.

The upward shift of high-energy optical N vibrations in HfNCl results from so called lanthanide contraction effect, which refers that the radii of the elements across the lanthanide series are unexpected decrease greatly, due to poorly shielding 4f orbitals and strongly attracting to outer shell electrons.[41] Indeed, the radius of Zr in fifth period is 1.75 Å while the radius of Hf in sixth period is also 1.75 Å, indicating a lanthanide elemental radius shrinking.[42] The bond length M-N ($d_2$ and $d_3$ in Table 1) also shows that $d_{Hf-N}$ is indeed shorter than $d_{Zr-N}$. We also perform the crystal orbital Hamilton population (COHP) calculation to demonstrate the chemical M-N bonding analysis for these two compounds, as shown in Fig. 5. The energy integrated COHP (ICOHP) values, which provide an estimate of the bond energy (the more negative value, the stronger the bonding strength),[43,44] are also shown in the figure. It can be seen that both $d_2$ and $d_3$ are indeed stronger in Hf compound than that of Zr compound. The stronger Hf-N bondings, therefore, cause the high-energy N vibrations shift upward in HfNCl.

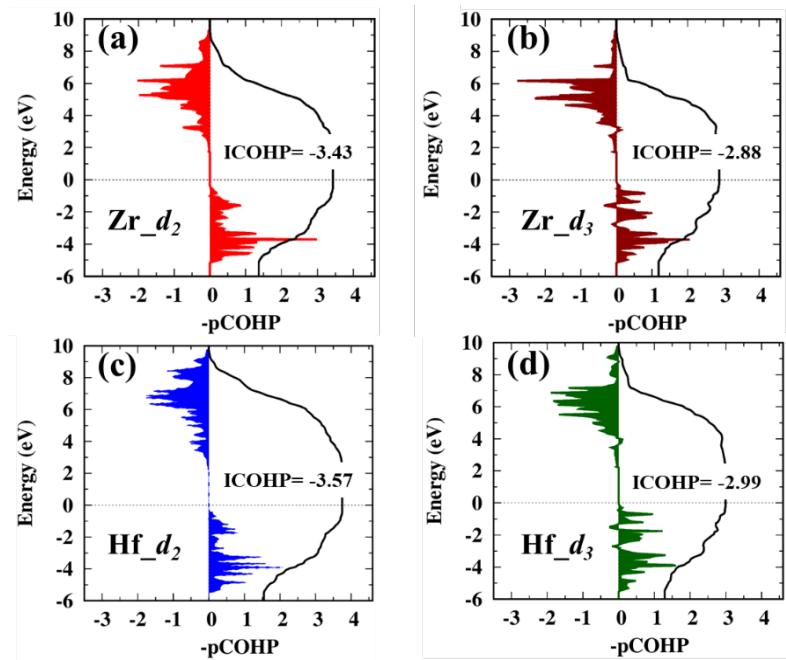

Fig. 5 The crystal orbital Hamilton population COHP (solid fill) and integrated COHP (black lines) for two different M-N bonds in ZrNCl (a, b) and in HfNCl (c, d). The –ICOHPs of $d_2$ are 3.43 and 3.75 eV, while the -ICOHPs of $d_3$ are 2.88 and 2.99 eV for Zr and Hf compounds. The black dashed line is Fermi level.

Now let's revisit the mass effect in the lattice thermal conductivity, using a simple one-dimensional harmonic two-atom chain, as shown in Fig. 6(a). Usually, the heavy mass atom with large atomic radii, generates a weak harmonic force with adjacent atoms and induces low acoustic phonons as well as low group velocity (see Fig. 6(c)), and therefore small lattice thermal conductivity. However, this only considers the group velocity effect on the thermal conductivity without considering the scattering rates change (Eq.1). This usually works because the optical phonon shifts downward as the interatomic interaction decreases due to longer bond length with heavier atom (larger atomic radius), see Fig. 6(b). However, if the replacing atom has lanthanide contraction effect, i.e., heavier mass but similar radius, then the newly formed bond may be even stronger than the previous one due to poorly shielding 4$f$ orbitals of heavy element. This will make the optical phonon branch shift upward and therefore broadens the phonon gap, as demonstrated in Fig. 6(c). In this case, though the heavy element causes the low group velocity and tends to reduce $\kappa_{latt}$, the broad phonon gap will suppress the phonon-phonon scattering rates and tend to enhance $\kappa_{latt}$. Therefore, there is competition between mass effect and lanthanide contraction effect and more attention should be paid to if using heavier lanthanide element to replace the light element.

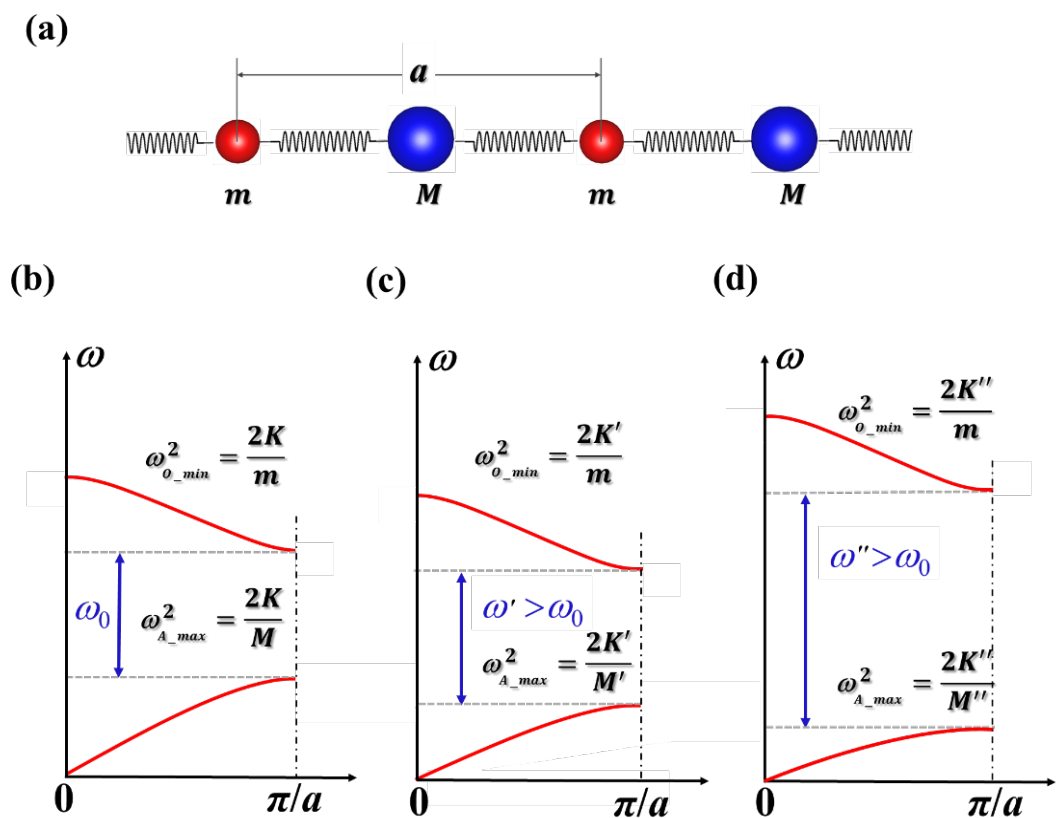

Fig. 6 Schematic change of phonon dispersion with heavy element replacement. (a) one-dimensional harmonic two-atom chain (lattice constant $a$, spring constant K, mass $M > m$) and its related phonon dispersion (b). (c) phonon dispersion after heavy element replacement

without lanthanide contraction effect ($M' > M$, $K' < K$). (d) phonon dispersion with the replacement of heavy element with lanthanide contraction effect ($M'' > M, K'' > K$). The $\omega_{O\_min}^2$ is the minimum phonon energy for the optic mode at zone boundary and $\omega_{A\_max}^2$ is the maximum energy for acoustic mode at zone boundary.

The lanthanide contraction effect not only induces abnormal thermal conductivity behavior in ZrNCl and HfNCl, it also brings unusual elastic properties in these two materials. Table 2 lists the elastic constants calculated from the finite difference method implemented in VASP[45] and the acoustic group velocity obtained by linearly fitting the phonon dispersion curve in Fig. 2 within q=0.05 range near zone center. The elastic constant C is proportional to the square of group velocity and density[46] and usually the materials with large elastic constant possess high thermal conductivity. The density effect is paid less attention because group velocity is more sensitive and has more impact on the elasticity than the density. The density effect is limited, for example, heavy atom replacement usually induces large volume and hence small change of density. However, due to the lanthanide contraction effect, after Hf replacement of Zr, the density increases dramatically by ~70% (see Table 2), this causes HfNCl possessing higher thermal conductivity but smaller group velocity than those of ZrNCl.

Table 2 Elastic constants $C_{ij}$ (in GPa), Young's modulus (in GPa), density $\rho$ (in gcm$^{-3}$) and velocities $v$ (in km/s) of acoustic phonon branches of ZrNCl and HfNCl along Γ-L direction in rhombohedral phase.

|  | $C_{11}$ | $C_{33}$ | $C_{44}$ | $C_{12}$ | $C_{13}$ | $E$ | $\rho$ | $v_{LA}$ | $v_{TA1}$ | $v_{TA2}$ |
|---|---|---|---|---|---|---|---|---|---|---|
| ZrNCl | 168.7 | 48.8 | 11.5 | 83.9 | 18.1 | 69.0 | 4.4 | 6.6 | 2.0 | 3.3 |
| HfNCl | 186.6 | 50.8 | 11.5 | 90.3 | 19.3 | 71.8 | 7.4 | 5.3 | 1.6 | 2.6 |

IV. CONCLUSIONS

In summary, we perform first-principles calculations to investigate the phonon transport of layered transition metal nitride halides MNCl (M= Zr, Hf) and find larger lattice thermal conductivity in heavy element HfNCl than ZrNCl. This behavior is mainly due to the lanthanide contraction effect in which Hf has much heavier mass than Zr but with similar radius. Though heavy mass of Hf induces low group velocity and tends to reduce the thermal conductivity, the stronger bonds of Hf-N makes high-energy optical phonons shift upward and broaden the phonon gap. This broadening gap reduces the scattering phase space and suppresses the three-phonon scattering channels of (aoo) and (ooo) processes, which enhances the lattice thermal conductivity of HfNCl significantly, compared with ZrNCl. Our work shows that the simple and effective strategy of heavy mass replacement to reduce the lattice thermal conductivity should be paid more attention, especially when the element in lanthanide series is involved. This also provides an alternative approach to reduce thermal transport properties in the materials with light lanthanide elements for designing effective thermoelectric materials.

## ACKNOWLEDGMENTS

This work is supported by the National Science Foundation of China (Grant No. 11572040), the National Materials Genome Project (2016YFB0700600) and the Technological Innovation Project of Beijing Institute of Technology. Theoretical calculations were performed using resources of the National Supercomputer Center in Guangzhou, which is supported by the Special Program for Applied Research on Super Computation of the NSFC-Guangdong Joint Fund (the second phase) under Grant No. U1501501. X.W. acknowledges the National Natural Science Foundation of China (Grant No. 11604011). X.X. Yu would like to thank Lina Yang for the discussion.